\documentclass[12pt,preprint,authoryear]{aastex}

\newcommand{\Ebv}{E_{\rm\scriptscriptstyle B-V}}
\newcommand{\Av}{A_{\rm\scriptscriptstyle V}}

\newcommand{\taus}{\tau_{\rm s}}
\begin{document}
\title{ KH\,15D: a star eclipsed by a large-scale dusty vortex? }
\shorttitle{ Dusty vortex around KH\,15D }
\shortauthors{Barge and Viton}
\author{P.\,Barge and M.\,Viton}
\affil{{\rm Laboratoire d'Astrophysique de Marseille, Traverse du Siphon B.P.8,
13376 Marseille cedex 12, France; Pierre.Barge@oamp.fr, Maurice.Viton@oamp.fr
}}

\begin{abstract}
We propose that the large photometric variations of KH\,15D are due to an eclipsing swarm of solid particles, trapped in a giant gaseous vortex rotating at $\sim 0.2$ AU from the star. The efficiency of the capture-in-vortex mechanism easily explains the observed large optical depth. The weaker opacity at mid-eclipse is consistent with a size segregation and a slow concentration of the particles toward the center of the vortex. This dusty structure must extend over $\sim 1/3$ of an orbit to account for the long eclipse duration. The estimated size of the trapped particles is found to range from 1\,cm to 10\,cm, consistent with the grey extinction of the star. The observations of KH\,15D support the idea that giant vortices can grow in circumstellar disks and play a central role in planet formation. 
\end{abstract}


\section{ Introduction }
KH\,15D is a pre-main sequence (PMS) K7 star located in the young star cluster, NGC\,2264. Its large photometric variations are periodic and possibly interpreted in terms of eclipses by an opaque feature orbiting inside a circumstellar disk seen nearly edge-on \citep{Her02}. It is likely connected to the UX\,Ori star family, characterized by large photometric and polarimetric variabilities \citep{Nal99}, but sets apart from them with two important differences: no infrared excess in the disk and no significant color changes during the eclipses, both indicating particle sizes much larger than optical wavelengths. As the $VRI$ colours of the star are consistent with its spectral type \citep{Ham01}, the selective absorption is $\Av < 0.2$, a value indeed in agreement with the foreground interstellar reddening, $\Ebv \simeq 0.06-0.07$, of stars in NGC\,2264 \citep{Perez87, Sung97}. 
On the other hand the grey absorption, $A_0$, during the out-of-eclipse observations can be estimated indirectly using a well known relation \citep{BEM78} between the visual brightness parameter and the intrinsic colour $(V-R)_0$. Using the absolute visual magnitude deduced from the distance modulus of the cluster, this relation provides a value of the stellar radius consistent with that derived by \citet{Ham01} and with the PMS status only if the grey absorption is $\sim 0.6 - 0.7$ magnitude. 

The five eclipses observed since the discovery of KH\,15D \citep{Kear98} provide us with the main characteristics of the occultation: a relative depth $\sim 0.95$, an approximate duration of 18 days and a period of 48.36 days. The absorbing cloud is deduced to orbit at $\sim 0.2$ AU from the star and to extend over ~1/3 of an orbit \citep{Ham01}. Such an occultation completely differs from a standard planetary transit and also displays other important peculiarities: (i) ingress and egress are remarkably steep, indicating an occulting clump with a very sharp edge; (ii) the minimum is not a flat land but contains a central bump with small scale features and a reverse peak close to mid-eclipse; (iii) when comparing two successive occultations, the reverse peak seems to shift from one side of the mid-eclipse to the other; (iv) since the first series of observations, the occultation has clearly widened and deepened in secular fashion. 

The various models suggested by \citet{Her02} to explain these observations involve a companion, planet or brown dwarf, able to $``$shepherd" gravitationally an extended swarm of solid particles (warped density wave or ring arc) which is responsible for the eclipses. Possibly, these models could account for a number of the observed characteristics but, so far, none of them has been developped far enough to reproduce the observed light-curve and no observation has confirmed the presence of a $``$shepherd". 

In this letter, we propose a new model in which KH\,15D is periodically occulted by a dusty anticyclonic vortex persisting for a long time in a gas disk surrounding the star. The particles are confined inside the vortex by the gas friction and no gravitational assistance is needed to keep them together into a clump. The swarm of the trapped particles is organized by the gas dynamics with an absorption profile that pretty well explains most properties of the observed light-curve. The gaseous vortex associated to this dust swarm covers a wide azimuthal extent (see Fig.1) which seems quite puzzling to explain but looks like those produced in recent numerical simulations \citep {KB03}. Our model is based on a scenario proposed by Barge and Sommeria (1994,1995; thereafter BS94,BS95) in which the formation of planets begins inside persistent gaseous vortices; it relies on a number of recent works on the formation and evolution of vortices in circumstellar disks \citep {Li01, KB03}.

\section{ The proposed model }

Our model starts from a circumstellar disk of gas and solid particles in which the mechanisms of planetary formation are still at work.
The gas is assumed to spread into a flared disk following a simple hydrostatic equilibrium in the vertical direction with a scale height $H=C_{\rm S} /\Omega$, where $C_{\rm S}$ is the sound speed and $\Omega$ is the Keplerian orbital frequency. Protoplanetary disks are likely turbulent during a span of their live and could host large scale and long-lived vortices growing from turbulence by analogy with what happens in two-dimensional fluid dynamics, in which organized structures are known to emerge from random turbulence in rotating shear flows. Vortices spinning like the shear flow are robust and merge one another while those with the opposite sign are laminated by the shear (BS95 and references therein). However, such vortices may also result from some specific instabilities as for example: (i) the Rossby Wave instability investigated by \citet {Li01} which requires a strong local maximum in entropy and density; (ii) a global baroclinic instability described recently by \citet {KB03} and which arises from a natural radial stratification of the gas flow. 

A number of numerical simulations show that large scale vortices can grow and survive for many rotation periods and look like either elongated structures stretched by compressibility effects \citep {GoL00} or a single vortex dominating the whole disk \citep {KB03}. Highly elongated vortices are found also in the inner regions of MHD accretion disks (Tagger and Pellat 1998, and private communication) or as exact solutions of the incompressible Euler equation with different aspect ratios \citep{Cha00}. On the other hand, the radial extent of vortices cannot exceed the thickness of the disk ($R<H$) since the velocity of the vortex, $R\Omega$, must be less than the sound speed $C_{\rm S}=H\Omega$ (BS95). Larger vortices would be destroyed by energy losses due to sound waves and density waves.
The solid particles embedded in the gas disk are submitted to a friction drag depending on the mean-free-path $\lambda$ of the gas molecules relative to the particle size (or radius) $s$. In the inner disk region ($r\leq 0.2$AU) and for particles larger than a millimeter, the mean-free-path is less than size and the drag, caused by particle wake, reaches the Stokes regime (in the limit of particle Reynolds numbers $Re_{\rm p} < 1$) with a stopping-time \citep{Cha00}:
\begin{equation}
T_s = {{8s^2\rho_0} \over{9\sigma_{\rm gas} \Omega \lambda}}~~, 
\end{equation}
where $\sigma_{\rm gas}$ is the gas surface density and $\rho_0 \simeq 2{\rm g/cm}^3$ is the density of material particles are made of. In contrast, the Epstein regime is reached for smaller particle size $s<9\lambda /4$. The particles are only submitted to star attraction and gas drag; their dynamical evolution depends on the single non-dimensioned friction parameter $\taus = \Omega T_{\rm s}$: 
(i) the lightest particles ($\taus \ll 1$) come at rest rapidly with the gas and are driven by the flow; (ii) the heaviest particles ($\taus \gg  1$) are nearly unaffected by the gas motion and keep a quasi-Keplerian motion. 

The capture of particles by a gaseous vortex has been explored first by BS94 and BS95 using a simple model in which the velocity field is made up of concentric epicycles inside the vortex, while it matches a Keplerian flow at large distances. They found that such an anticyclonic vortex can capture and concentrate dust particles very efficiently: (i) light particles penetrate into the vortex and stop on streamlines close to the edge where they slowly shift toward the core; (iii) optimal particles, with $\taus \sim  1$, sink deeply into the vortex and reach core streamlines. 
These results were confirmed for other velocity fields \citep{Cha00, Fuen02}. The vortex induces a segregation of the trapped particles (or a size sorting if the particles have the same composition) following the value of their friction parameter; this is illustrated in a number of numerical simulations \citep{Fuen02}. 

The ultimate reason for this dynamical behaviour lies in the sign of the vortex rotation. Indeed, in a reference frame rotating with the vortex center, the Coriolis force can overcome the centrifugal force and pushes the particles toward the core if the vortex is an anticyclone, whereas both forces are conspiring to eject the particles for a cyclone. This capture-in-vortex mechanism is a very efficient one and results in strong density enhancements inside the vortex, by at least two order of magnitude in $\sim 200$ rotation periods. The capture rate is estimated under the assumption that the particles are continuously renewed near the vortex orbit due to the inward drift under the systematic headwind drag (Weidenschilling 1977, BS95). One obvious and important consequence of this density enhancement is that, inside a vortex, particle growth is made easier and will couple with confinement and segregation. On the other hand, dust is depleted from the region inside the vortex orbit as the particles either are feeding the vortex or are falling to the star under the systematic drift.

Inside the vortex, the trapped particles are also submitted to a background small scale turbulence which makes them diffuse and tends to reduce their global concentration. \citet{Cha00} investigated this question in terms of a diffusion equation in an idealized circular vortex and derived a time dependant solution for the surface density inside the vortex: 
\begin{equation}
\sigma_{\rm d}  \propto \frac{1}{l_{\rm d}^2 (1-k^2)}~~ 
{\exp\left[-\frac{(r-k r_0)^2}{l_{\rm d}^2 (1-k^2)}\right]}~~,
\end{equation}
the initial state being a delta function centered at $r_0$; $l_{\rm d}$ is the diffusion scale length and $k = \exp(-t/T_{\rm capt})$ in which $T_{\rm capt}$ is the characteristic time for a particle to reach the center of the vortex. In the case of light particles and very elongated vortices $T_{\rm capt} \simeq 2q/(3\Omega\taus)$ and $l_d\sim \sqrt{\alpha_{\rm v}/q\taus}~R_{\rm v}$ where $R_{\rm v}$ is the vortex radius and $q$ its aspect ratio; $\alpha_{\rm v}$ is the non-dimensioned parameter measuring the small scale turbulence efficiency inside the vortex.

In order to estimate the size of the trapped particles, we will choose a standard model of nebula, the minimum mass solar nebula, in which the surface densities (both for gas and particles) and the temperature are the decreasing power laws $r^{-3/2}$ and $r^{-1/2}$, respectively; at 1 AU the densities are set to 1700 ${\rm g\,cm}^{-2}$ for the gas and 20 ${\rm g\,cm}^{-2}$ for the particles, whereas the temperature is assumed to be 280\,K. At 0.2 AU from the star and with our numerical values, the optimal size for particle capture is $s_{\rm opt} \simeq 31\sqrt{\taus}(r/1{\rm AU})^{5/8}\simeq 11$\,cm, i.e. pebble size.
Lighter non optimal particles with $\taus<1$ can be also captured by the vortex but remain trapped at the periphery of the vortex. 
These particles remain in the Stokes regime, at 0.2 AU from the star, as far as their size is larger than the critical size $s_{\rm c} = 9\lambda /4\simeq 0.3$ mm.

\section{ Fitting the light-curve }

We assume the disk is seen nearly edge on, under an inclination $i$ less than the flaring angle of the gas disk. $i$ must be small enough for the line of sight to cross the vortex in its vertical extent, but also large enough to avoid the prohibitive optical depths near the mid-plane layers. Of course, this inclination has to be consistent with the out-of-eclipse grey extinction $A_0 \sim 0.7$ magnitude deduced in Sect.1. In its motion around the star, the vortex periodically crosses the line of sight and the optical depth $\tau$ is a function of time varying from $A_0$ (out of eclipse) to $A_0 + 3.2$ during the eclipses. Assuming the absorption is due to spherical particles with sizes larger than a millimeter, the opacity $\kappa$ reduces to $\pi s^2 n$ and the optical depth $\tau = \int \kappa \,dl$ is the familiar dust column density, where $n$ is the local number density along the line of sight. 

During an occultation $\tau$ is found to increase inside the vortex, from center to edge, then to fall down very steeply to reach the out-of-vortex level. This $``$opacity-curve" has a peculiar shape that our model can easily reproduce.

First, one can guess that a circular blob of matter in which the opacity $\kappa (r)$ is radially increasing can conveniently mimic the optical depth profile. Then, we noticed that a simple linear dependance, $\kappa \propto 1 + a~r$, with a cut off at the outer boundary permits to fit pretty well the data (Fig. 2). 
In our fit the optical depth has been computed following two steps: (i) an integration of $\kappa $ along a path crossing an equivalent circular vortex (ECV) of radius $H$ and located at a distance $r_{\rm orb}$ from the star; (ii) a circular anamorphosis by a factor $q\simeq 17$ (the result of an integration along a path at angle $u$ is reported to the  direction of observation at angle $q\times u$). Figure 1 illustrates the ECV model with a sketch of the integration path and the density contours of the vortex based on the linear opacity law and the appropriate anamorphosis.

This absorption profile corresponds to a shell-like distribution of the solid particles inside the vortex, reminiscent of the spherical structures observed in H{\sc ii} regions. In our model such a density profile originates in the dynamics and the segregation of particles trapped in a vortex as described above. 

Surprisingly, a simple cut off at the vortex outer boundary can easily mimic the very steep fall during ingress or egress, consistent with an occultation by a sharp edge as proposed by \citet{Her02}. In fact, a cut off cannot receive true physical justification and the resulting profile cannot connect smoothly the out-of-eclipse level. A more realistic fit of ingress and egress directly arises from our model. Indeed, light sub-optimal particles are preferentially captured in the outer part of the vortex where they accumulate and slowly diffuse toward the core. The fit is realized thanks to the Gaussian density profile presented in Eq.\,2 with the following values of the parameters: the center of the Gaussian is close to the vortex boundary (at a distance from the core $\sim R_{\rm v} = q\,H$); the $1/e$ half-width of the Gaussian is given by $l_{\rm d}\sqrt{(1-k^2)}\simeq 0.07 R_{\rm v}$. This is possible using the following assumptions: (1) the particles are renewed by capture every $\sim 10$ rotations of the vortex and have a friction parameter $\taus\simeq 0.01$, consistent with a capture at the vortex periphery and a slow drift toward the core in a time scale $T_{\rm capt} \simeq 2q/(3\Omega \taus)$; (2) the parameter measuring the turbulence inside the vortex is $\alpha_{\rm v} \simeq 0.01$. As a result, the correponding size of particles confined in the outer regions is of the order of a centimeter.

\section{ Discussion and conclusion}

The proposed model can account for the main characteristics of KH\,15D's observations. It is able to reproduce the light-curve and permits to fit pretty well the data.

(1) The period and duration of the eclipses may be reproduced by the rotation around the star of a giant swarm of solid particles, trapped in a persistent gaseous vortex.

(2) During totality, the rise up of the flux can result from a lower opacity of the central regions, possibly coming from a particle size segregation (the larger the particle, the deeper and the faster the confinement inside the vortex).

(3) The estimated size of the trapped particles ranges from 1 to 10 centimeters, consistent with the grey absorption observed in KH\,15D. The presence and growth of bigger particles is very likely due to the high densities and the low relative velocities. This is in agreement with the expected predominance of large grains in PMS disks which seems also required to explain sub-millimeter and millimeter disk observations \citep{Nal99}. 

(4) The shift of the reverse peak observed in successive transits could be explained by gas rotation inside the vortex, the inner zone rotating at half the orbital frequency similarly to what happens in smaller vortices \citep{Cha00}.

We can also speculate that: (i) $``$secular" changes of the light-curve could result from a dynamical evolution of the swarm of particles; (ii) the wavy behaviour at egress could correspond to a turbulent wake of the vortex.

One of the main assumption in our model is that the disk is seen nearly edge-on, at an angle $i\sim 2-3^\circ$, and with an optical depth $A_0\sim 0.6-0.7$ magnitude during the out-of-eclipse periods. This is a possible situation if the dust particles have grown and settled down to the mid-plane forming a $\sl flat$ sub-disk dominated by large grains. The predominance of large particles in KH\,15D is, indeed, completely consistent with our results. However, we want to stress that no numerical simulation is, nowadays, able to reproduce the huge azimutal extent of the vortex required to fit the observations. Further hydrodynamical simulations and theoretical modeling of the particle dynamics are necessary to confirm and improve our model.

Finally, KH\,15D appears a good target to test planetary formation models. This is presently the case with the scenario in which vortices can persist in a gas disk acting as traps for dust (protected against losses into the star) and as wombs for planetesimals. Of course, further observations of this object are required to constrain more firmly the model. The possibility to observe large scale structures in circumstellar disks has been recently investigated by \citet{Wol02}, extrapolating the capabilities of the future giant interferometer ALMA. KH\,15D and other similar objects are good opportunities for the future space missions like CoRoT, Kepler and Eddington.

A further piece of information was recently provided by \citet{Ham03} 
using high resolution UVES spectra which confirmed that KH\,15D is a weak-lined T Tauri star  surrounded by an accretion disk with, possibly, a collimated bipolar jet. Such configuration could favor the accretion-ejection model of \citet{TaPe98} in which persistent vortices can form  in the inner region of magnetized disks.

While submitting this letter we learned that H.\,Klahr also claims the existence of a vortex in the disk of KH\,15D to explain the peculiar light-curve of this object. His work seems complementary to ours since he is focusing on the way a giant vortex forms and persists in hydrodynamical simulations, and not on the trapping of solid particles in gaseous structures.

	We wish to thank W.\,Herbst and C.M.\,Hamilton for communicating us the KH\,15D data plotted in Fig.\,2. We are also grateful to 
A.\,Llebaria for his help in handling the data.

\clearpage

\begin{figure}[h]
\plotone{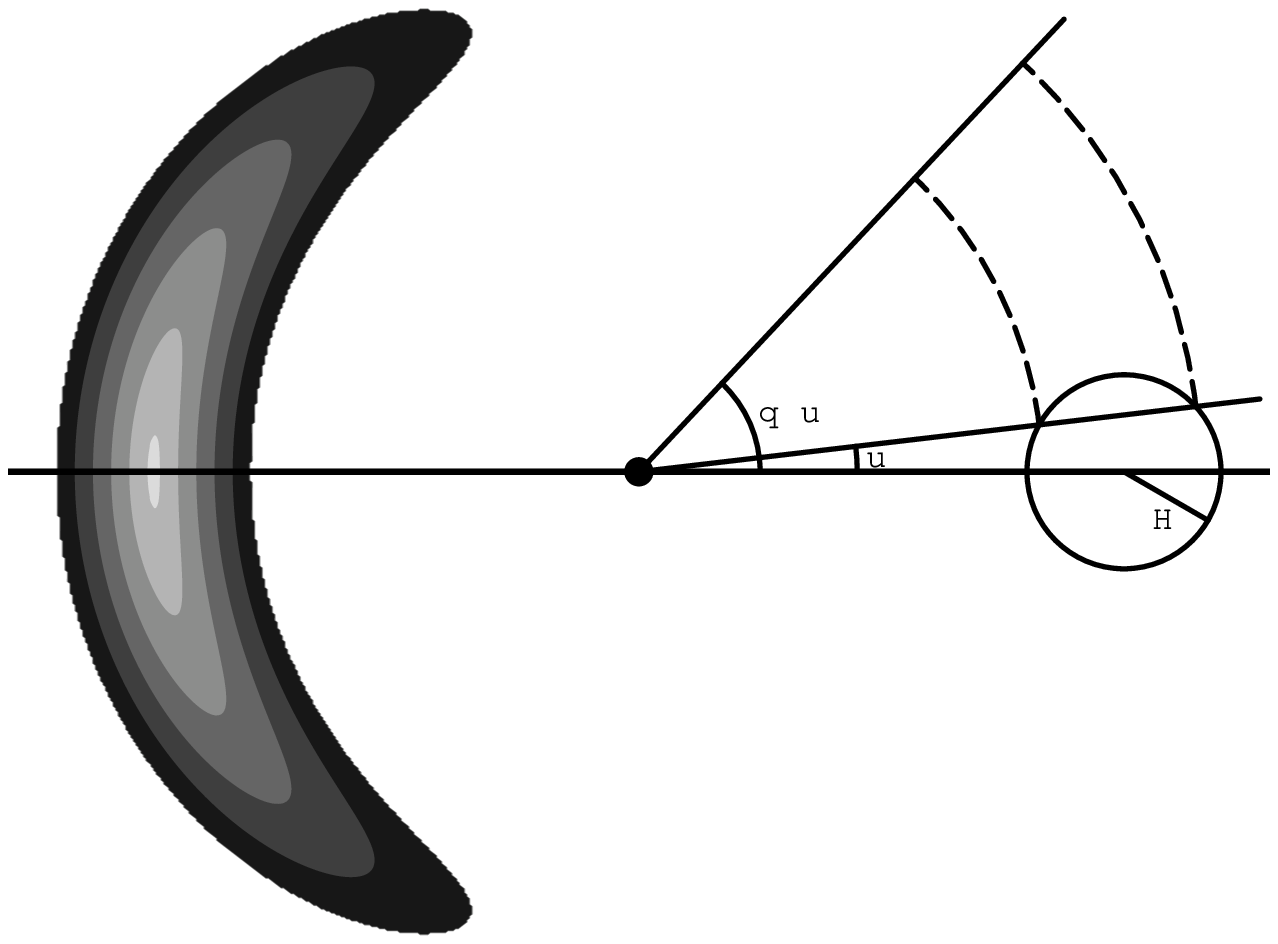}
\caption{The banana-like dusty vortex used to model the eclipses of KH\,15D. {\em Right:} Sketch of the equivalent circular vortex (ECV) of radius $H$ centered at a distance $r_{\rm orb}$ from the star. The optical depth obtained by integrating the opacity along a path crossing the ECV at angle $u$ is reported to the direction of observation at angle $q\times u$. {\em Left:} Approximate density contours (0.15 to 0.9 in steps of 0.15) corresponding to the {\em linear} opacity law (see text), a crude model surprisingly representative of the observed extinction curve. Note that the ratio $H/r_{\rm orb} =0.2$ used in this figure is much larger than actual. } 
\end{figure}
\begin{figure}[h]
\plotone{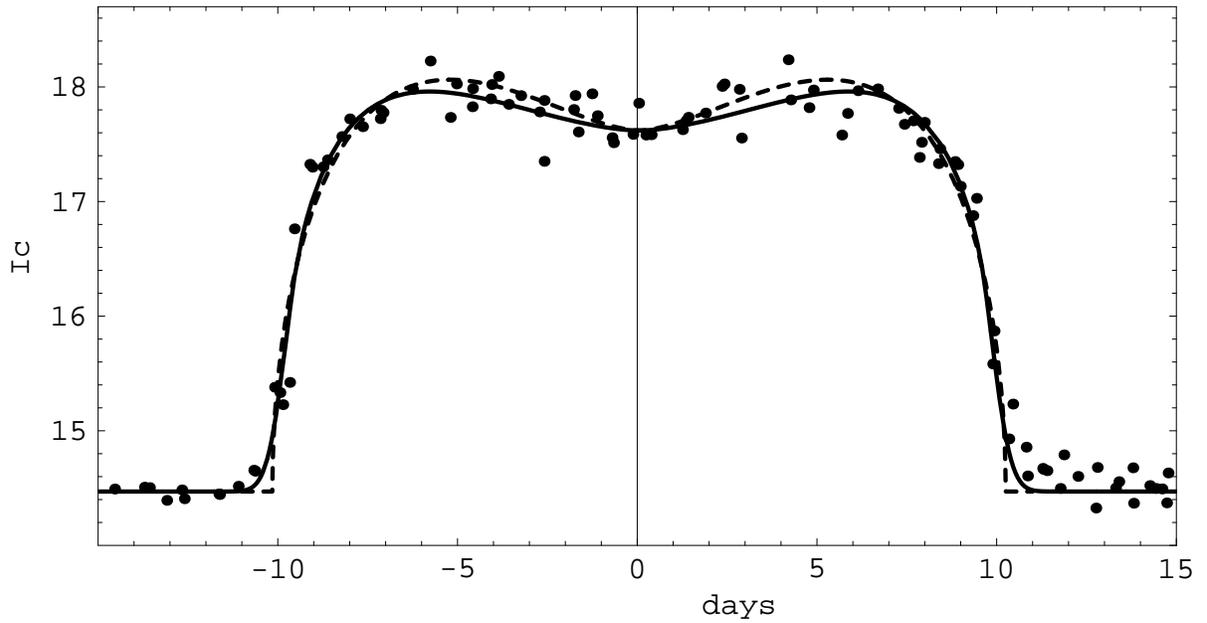}
\caption{Eclipses of KH\,15D in $I_{\rm \scriptscriptstyle C}$-band magnitudes, phased with the period of 48.36 days. The plotted points are weighted means per observation night from the $2001-2002$ observation campaign (W. Herbst and C. Hamilton, private communication). {\it Dashed} line is a fit obtained with the linear approximation $\kappa \propto 1 + a~r$ and $a\sim  10$. {\it Solid} line is the best fit obtained assuming $\kappa \propto b^r$ ($b \sim 5$) if $r<H$, and $\kappa$ given by the Gaussian model described in the text if $r \ge H$.}
\end{figure}   
\end{document}